\documentclass[prb,aps,twocolumn,home,floats]{revtex4}
\usepackage{graphics}
\begin{document}
 \title {{\rm\small\hfill  (submitted to Surf. Sci.)}\\
 The \boldmath Pd(100)-$(\sqrt{5} \times \sqrt{5})R27^o$-O surface
 oxide revisited}

 \author{M. Todorova${}^1$, E. Lundgren${}^2$, V. Blum${}^3$, A.
 Mikkelsen${}^2$, S. Gray${}^2$,\\
 J. Gustafson${}^2$, M. Borg${}^2$, J. Rogal${}^1$, K. Reuter${}^1$,
 J.N. Andersen${}^2$, and M. Scheffler${}^1$}

 \address{${}^1$ Fritz-Haber-Institut der Max-Planck-Gesellschaft, Faradayweg
 4-6, D-14195 Berlin, Germany}

 \address{${}^2$ Department of Synchrotron Radiation Research, Institute of
 Physics, Lund University, Box 118, SE-221 00 Lund, Sweden}

 \address{${}^3$ National Renewable Energy Laboratory, Golden, Colorado 80401,
 USA}

 \date{\today}

 \begin{abstract}
 Combining high-resolution core-level spectroscopy (HRCLS),
 scanning tunneling microscopy (STM) and density-functional theory
 (DFT) calculations we reanalyze the Pd(100)-$(\sqrt{5} \times
 \sqrt{5})R27^o$-O surface oxide phase. We find that the prevalent
 structural model, a rumpled PdO(001) film suggested by previous
 low energy electron diffraction (LEED) work (M. Saidy {\em et al.},
 Surf. Sci. {\bf 494}, L799 (2001)), is incompatible with all three
 employed methods. Instead, we suggest the two-dimensional film to
 consist of a strained PdO(101) layer on top of Pd(100). LEED 
 intensity calculations show that this model is compatible with
 the experimental data of Saidy {\em et al.}\\
 \end{abstract}

 \maketitle

 \section{Introduction}

 Due to their importance for catalysis and corrosion, oxidation
 processes at transition metal (TM) surfaces have long received
 significant attention both in fundamental and applied research.
 Roughly divided into pure on-surface (dissociative) adsorption,
 surface-oxide formation, and oxide film growth, particularly the
 second step in the oxidation sequence is poorly understood
 on a microscopic level. Not least, this is due to the fact that an
 atomic-scale investigation of TM surfaces on the verge of oxide
 formation still poses a significant challenge for the otherwise
 well-developed machinery of ultra-high vacuum (UHV) surface
 science: The rather high oxygen partial pressures and elevated
 temperatures required to initiate oxide nucleation, a low degree
 of order at the surface and highly complex, large unit-cell
 geometries name but a few of the problems encountered on the route
 towards a microscopic characterization of surface oxides.

 The vast structural parameter space connected with surface oxides
 is often prohibitive for an analysis based on one technique
 alone, in particular for the exhaustive searches required in
 diffraction-based structure determination techniques like low
 energy electron diffraction (LEED). Fortunately, over the last
 10 years it became clear that the joint effort of experimental
 works and theory (in particular when using first-principles
 methods) is synergetic and most valueable, if not crucial, and
 allows a convincing identification and characterization of
 complex, novel structures at surfaces
 \cite{schmalz91,burchhardt95,stampfl96,lee00,lundgren02}.
 Such a multi-method approach is also adopted in the present work,
 where high-resolution core-level spectroscopy (HRCLS), scanning
 tunneling microscopy (STM), and density-functional theory (DFT)
 are employed addressing the Pd(100)-$(\sqrt{5} \times
 \sqrt{5})R27^o$-O surface oxide phase (coined $\sqrt{5}$-phase
 in the following for brevity), which for $T > 400$\,K concludes
 the series of ordered phases observed on Pd(100) before
 three-dimensional cluster growth sets in \cite{zheng02}.

 A previous tensor LEED analysis suggested the $\sqrt{5}$-phase to
 correspond essentially to a rumpled PdO(001) plane on top of
 Pd(100) \cite{vu94,saidy01}. In the following we will show that
 this assignment can not be reconciled with either HRCLS, STM, or
 DFT, the methods employed in the present study. Instead, we
 propose a new model, consisting of a strained PdO(101) layer on
 Pd(100). This is in agreement with all experimental data and
 energetically more stable than the previous model. Moreover,
 we performed preliminary LEED calculations to show that this new arrangement
 can also be reconciled with the previous experimental LEED data
 of Saidy {\em et al.} \cite{saidy01}. Interestingly, the (101)
 orientation does not correspond to a preferred growth direction
 \cite{mcbride91} or a low-energy surface of crystalline PdO,
 suggesting that the film-substrate interaction may stabilize
 higher energy crystal faces. We argue that the atypical orientation
 and the resulting chemical properties of such thin oxide layers
 may be of interest in future applications.

 \section{Experimental and computational details}

 The STM measurements were performed in a UHV chamber with a base
 pressure below $1 \times 10^{-10}$\,mbar. The Pd(100) surface was
 cleaned by cycles of Ar${}^+$ sputtering, annealing and oxygen
 treatments keeping the sample at 900\,K in an oxygen pressure of
 $2 \times 10^{-8}$\,mbar followed by flashes to 1400\,K. The
 cleanliness of the Pd(100) surface was checked by Auger Electron
 Spectroscopy (AES); no contaminants such as C and O could be
 observed within the detection limits. The $\sqrt{5}$-phase was
 thereafter formed by exposing the Pd(100) surface to an oxygen
 pressure of $5 \times 10^{-6}$\,mbar for 300 seconds at
 $T = 600$\,K.

 The HRCLS measurements were conducted at the beam line I311
 \cite{nyholm01}at MAXII in Lund, Sweden. The cleaning procedure
 and preparation of the $\sqrt{5}$-phase was identical to that
 described above. The HRCL spectra were recorded at liquid nitrogen
 temperatures and at normal emission angle. The cleanliness of the
 Pd(100) surface was checked by monitoring the Pd $3d_{5/2}$, O
 $1s$, and C $1s$ core-levels, as well as the valence band region;
 again, no contaminants could be detected.

 The DFT calculations were performed within the Full-Potential
 Linear Augmented Plane Wave (FP-LAPW) scheme
 \cite{blaha99,kohler96,petersen00} using the generalized gradient
 approximation (GGA) \cite{perdew96} for the exchange-correlation
 functional. The $\sqrt{5}$ surface oxide was modeled in a
 supercell geometry, employing a symmetric slab consisting of five
 layers Pd(100) in the middle plus the various PdO layers described
 below on both sides. A vacuum region of $\approx$ 15 {\AA} ensures
 the decoupling of the surfaces of consecutive slabs. All atomic
 positions within the PdO and the outermost substrate layer were
 fully relaxed.

 The FP-LAPW basis set parameters are as follows:
 $R_{\rm{MT}}^{\rm{Pd}}=$1.8 bohr, $R_{\rm{MT}}^{\rm{O}}=$1.3 bohr,
 wave function expansion inside the muffin tins up to
 $l_{\rm{max}}^{\rm{wf}} = 12$, potential expansion up to
 $l_{\rm{max}}^{\rm{pot}} = 4$, and a $(4 \times 4 \times 1)$
 Monkhorst-Pack grid with 8 {\bf k}-points in the irreducible
 part of the Brillouin zone (28 {\bf k}-points in the full zone).
 The energy cutoff for the plane wave representation in the
 interstitial region between the muffin tin spheres was $E^{\rm
 max}_{\rm wf}$ = 20 Ry for the wave functions and $E^{\rm
 max}_{\rm pot}$ = 169 Ry for the potential. With this basis set
 very good agreement is obtained with the results of the structure
 determination for the $p(2 \times 2)$ phase on Pd(100) reported by
 a previous LEED study \cite{kolthoff96}, and we find the relative
 energetic stabilities of the various tested overlayer models
 converged to within $\pm 50$\,meV per O atom.

 This does, however, not comprise the uncertainty in the absolute
 binding energies, $E_{B}$. DFT, even within the GGA, is known to poorly
 describe gas phase oxygen and gives in particular the binding
 energy for molecular O$_2$ wrong by about 0.5\,eV per O atom
 \cite{pirovano99}. As this directly effects the obtained
 absolute binding energies for the various surface oxide phases,
 a possible workaround would be to determine the total energy
 of molecular oxygen, $E^{\rm tot}_{{\rm O}_2}$, not via gas phase
 calculations, but via \cite{li03}
 \begin{displaymath}
 1/2 E^{\rm tot}_{\rm O_2} \;\approx\; E^{\rm tot}_{\rm PdO, bulk} - E^{\rm
 tot}_{\rm Pd, bulk} + \Delta H_f({\rm 300 K}, {\rm 1 atm})
 \end{displaymath}
 i.e. employing an approximate equation for the PdO heat of formation, $\Delta
 H_f$, into which only the total energies of bulk PdO and Pd bulk,
 $E^{\rm tot}_{\rm PdO, bulk}$ and $E^{\rm tot}_{\rm Pd, bulk}$, enter. Using
 the experimental $\Delta H_f({\rm 300K}, {\rm 1 atm})$ one can thus arrive at
 $E^{\rm tot}_{\rm O_2}$ without having to resort to atomic calculations, though
 at the expense of discarding a completely first-principles type description.

 In the present work we will only compare the stability of
 various structural models all including the same number of
 oxygen atoms. Then, the difference between the standard
 computation of binding energies, i.e. utilizing
 gas-phase computed $E^{\rm tot}_{\rm O_2}$, and the aforedescribed
 procedure amounts only to a constant shift in the calculated
 binding energies. Employing $\Delta H_f^{\rm exp}({\rm 300 K}, {\rm 1 atm})
 = 0.88$\,eV \cite{CRC95}, this shift amounts to 0.43\,eV
 per O atom with a lower stability of the $\Delta H_f$-derived
 binding energies and not including zero-point vibrations. This indicates
 the sizable uncertainty in the absolute binding energy values and
 correspondingly dictates a cautious judgement on the endo- or exothermicity of
 a structure. In the following, we will always indicate binding energies
 obtained with gas-phase computed $E^{\rm tot}_{\rm O_2}$ (not including
 zero-point vibrations), which according to the above argument are likely to
 represent an upper limit to the real stability. Our main conclusion, the
 rebuttal of the prevalent $\sqrt{5}$-model, will, however, rather be based on
 relative energetic differences, which are fortunately enough much better
 defined.

 The surface core-level shift (SCLS), $\Delta_{\rm{SCLS}}$, is
 defined as the difference in energy which is needed to remove a
 core electron either from a surface or a bulk atom
 \cite{spanjaard85}. In the initial-state approximation the SCLS
 arises simply from the variation of the computed orbital
 eigenenergies before the excitation of the core electron. In
 final-state calculations, on the other hand, the SCLS involves an
 additional component due to the screening contribution from the
 valence electrons in response to the created core hole, obtained
 approximately via the Slater-Janak transition state approach of
 evaluating total energy differences using impurity type
 calculations as explained in detail in ref. \onlinecite{lizzit01}.
 In case of the Pd $3d$ SCLSs the bulk level position can be
 employed as a well defined reference level to align theoretical and
 experimental spectra. For the O $1s$ levels however, we note that it
 is not very practical to also use the Pd $3d$ bulk-level position to
 align the O $1s$ data, as both types of orbital eigenenergies
 exhibit different convergence behavior with respect to the employed
 basis set. Also the Fermi-level position as another reference level
 present in both experimental and theoretical spectra is not very
 practical: Particularly for experiments on systems with high oxygen
 loads like surface oxides band bending can not be excluded.

 Fortunately, we only need to rely on the existence (or non-existence) of a
 split O $1s$ spectrum in the present work. This difference in relative O $1s$
 level positions of various atoms within the same geometry is well defined and
 independent of the reference zero used. Just in order to present the
 theoretical and experimental data in the same plot (Fig. 2) we will therefore
 employ the simplest possible alignment approach given by equating the position
 of the lowest theoretical and experimental O $1s$ core-level position. We
 stress that this crude procedure does not enter our physical argument and is
 solely used for graphical purposes.

 \section{Results and discussion}

 \subsubsection{Shortcomings of the prevalent $\sqrt{5}$-LEED model}

 \begin{figure}
 \centering
 \scalebox{0.40}{\includegraphics{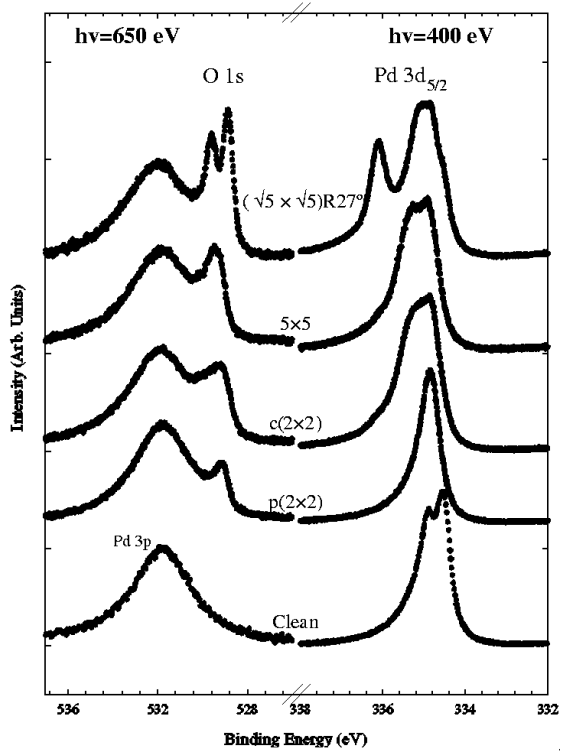}}
 \caption{Experimental HRCL spectra of the O $1s$ and the
 Pd $3d_{5/2}$ levels for the sequence of ordered structures
 that form on Pd(100) with increasing oxygen coverage. The
 photon energies were 650\,eV and 400\,eV respectively.}
 \label{fig1}
 \end{figure}

 Fig. 1 shows the development of the experimental HRCL spectra from
 the O $1s$ and Pd $3d_{5/2}$ levels for the sequence of ordered
 structures that form on the Pd(100) surface with increasing oxygen
 coverage. After the two known adsorption phases, $p(2 \times 2)$
 at $\theta = 0.25$\,ML and $c(2 \times 2)$ at $\theta = 0.50$\,ML
 \cite{kolthoff96,orent82,stuve84,chang88a,chang88b}, at first an
 intermediate $(5 \times 5)$ surface oxide forms, before finally
 the $\sqrt{5}$ is obtained, on which we concentrate in this work.
 Focusing first on the O $1s$ spectrum of this $\sqrt{5}$-phase,
 its most surprising feature is the existence of two sharp peaks in
 contrast to the single peak observed at all lower coverage
 structures. As is apparent from Fig. 1, a definite assignment of
 these peaks is hampered by the energetically very close lying Pd
 $3p$ levels. Still, we attribute these two peaks to emission from
 O $1s$ levels as no similar features are observed at the low
 binding-energy side of the Pd $3d$ spectrum. Changing the incident
 photon energy to vary the escape depth of the photoelectrons leads
 us to conclude that they originate from oxygen atoms close to the
 surface and at roughly the same depth.

 \begin{figure}
 \centering
 \scalebox{0.53}{\includegraphics{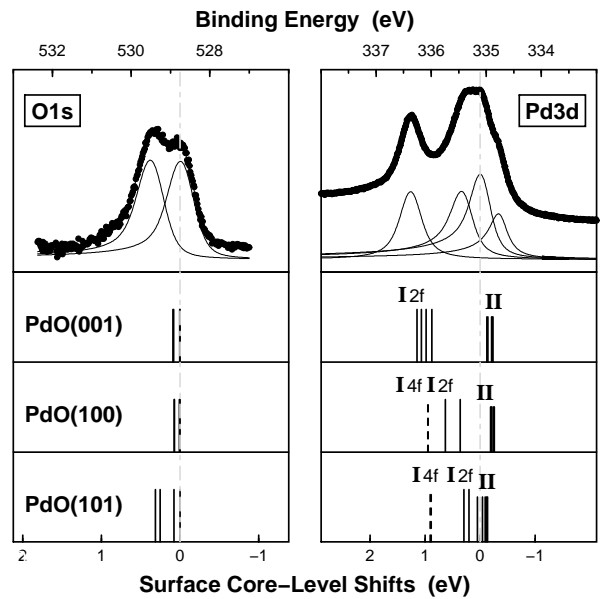}}
 \caption{Top panel: measured HRCL spectra of the O $1s$ and the
 Pd $3d_{5/2}$ levels from the Pd(100)-$(\sqrt{5} \times
 \sqrt{5})R27^o$-O phase at higher photon energies (900\,eV
 and 650\,eV for O $1s$ and Pd $3d$, respectively) and with the
 Pd $3p$ contribution of the clean surface removed.
 Bottom panel: Calculated final-state shifts for the three structural
 models shown in Fig. 3. For Pd $3d$ the bulk-level is employed
 to align theoretical and experimental spectra. For O $1s$ the
 lowest-energy theoretical peak is simply aligned to the lowest-energy
 experimental peak (see text). Note that only the PdO(101) layer on Pd(100)
 exhibits a split O $1s$ spectrum with two significantly shifted
 components. See Fig. 3 for the nomenclature used to describe the atoms
 from which the various theoretical Pd core-level shifts originate.}
 \label{fig2}
 \end{figure}

 To arrive at a rough estimate of the relative coverages, we
 approximately remove the contribution from the Pd $3p$ levels by
 subtracting the Pd $3p$ spectrum recorded from the clean surface.
 The corresponding HRCL spectra for the $\sqrt{5}$-phase are shown
 in Fig. 2, now recorded at higher photon energies of 900\,eV and
 650\,eV for the O $1s$ and Pd $3d$ levels, respectively. At the
 expense of a decreased resolution, these somewhat high energies
 are employed to avoid diffraction effects and thus permit a rough
 estimate of the relative coverages giving rise to the two peaks.
 We note that despite the decreased resolution, the two peaks are
 still clearly distinguishable in the original (not shown) and
 subtracted (Fig. 2) data. In both cases we obtain a considerable
 binding energy shift of $\approx 0.75$\,eV between both peaks and
 a ratio of about 1:1 for the two components. On the basis of this
 analysis of the experimental O $1s$ spectrum we would therefore
 anticipate at least two oxygen species at or close to the surface
 of the $\sqrt{5}$-phase in close to equal amounts.

Turning to the Pd $3d_{5/2}$ spectra in Fig. \ref{fig2}, at least
three oxygen-induced components are experimentally resolved, at
-0.32\,eV, +0.38\,eV and +1.30\,eV, where a positive SCLS
indicates a higher binding energy with respect to the reference Pd
bulk component. Using simple initial-state arguments, we expect an
increased oxygen coordination to yield a positive SCLS for the Pd
$3d_{5/2}$ level. Thus, the component shifted by +1.30\,eV could
be due to highly oxygen coordinated Pd atoms, agreeing with
previous results from the intermediate oxide structure on Pd(111)
\cite{lundgren02}, in which a very similarly shifted component was
found and assigned to Pd atoms fourfold coordinated to oxygen. The
component shifted by +0.38\,eV could correspondingly be due to Pd
atoms coordinated to less oxygen atoms (possibly two or three),
whereas the component shifted by -0.32\,eV likely originates from
Pd atoms at the interface between the Pd(100) substrate and the
thin oxidic film.

 \begin{figure}
 \centering
 \scalebox{0.44}{\includegraphics{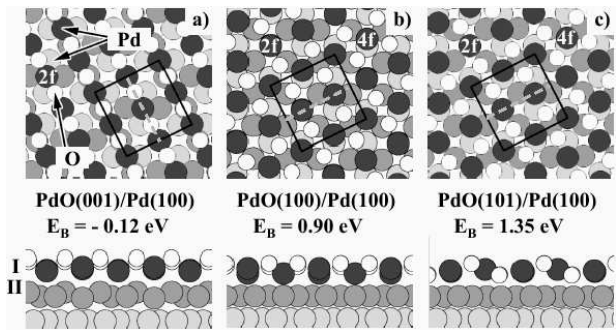}}
 \caption{Top- and side-view of the three structural models of the
 Pd(100)-$(\sqrt{5} \times \sqrt{5})R27^o$-O phase considered in
 the present work. All models assume an oxygen coverage of 0.8
 ML. a) PdO(001) layer on Pd(100) \cite{vu94,saidy01}, b) PdO(100)
 layer on Pd(100), and c) PdO(101) layer on Pd(100). The $\sqrt{5}$
 unit-cell is sketched in the top-views (solid line), while the
 dashed lines indicate the direction of atomic rows seen in the STM
 images. The DFT binding energy of the three models clearly reveals
 the PdO(101) layer on Pd(100) as the most favorable model. $2f$ and
 $4f$ denote two- and fourfold coordinated first-layer Pd atoms, and
 relate to the labels in Fig. 2 used to specify the atomic origin
 of the various computed core-level shifts.}
 \label{fig3}
 \end{figure}

 Trying to analyze the compatibility of the present HRCLS data with the
 prevalent structural model of the $\sqrt{5}$-phase suggested on the
 basis of the tensor LEED analysis (henceforth abreviated with ``LEED
 model'') \cite{vu94,saidy01}, cf. Fig. 3a, we used the published
 atomic positions of the LEED model as input to our DFT computations.
 Obtaining SCLS that did not resemble the experimental data at all,
 we initially proceeded by subjecting the LEED model to a complete structural
 relaxation. The resulting final-state SCLS after relaxation are shown in Fig. 2
 and are still difficult to reconcile with the experimental data:
 The large splitting of the O $1s$ spectrum is not reproduced and
 almost identical O $1s$ positions are obtained for all O atoms in
 the structure. Recalling that the LEED model essentially corresponds
 to a PdO(001) overlayer on Pd(100), in which all oxygen atoms are
 in principle equivalent, cf. Fig. \ref{fig3}a, this result is not
 surprising. The agreement in case of the Pd $3d$ SCLSs is not much better,
 obtaining computed shifts that are split into two distinct groups in
 contrast to the four component structure seen experimentally.

 Geometrically, the structural relaxation in DFT almost completely removes
 most of the strong rumpling introduced in the LEED study to fit the
 experimental $I(E)$-curves, and the substrate/oxide interface
 smoothes out at a large interface distance indicating a very weak
 coupling. In the end, the absolute binding energy per O atom, $E_B$,
 of the LEED model was (with the caveat given in Section II) still
 found to be slightly endothermic with respect to molecular oxygen
 ($E_B = -0.12$\,eV/ O atom), adding to our doubts of this prevalent
 structural model of the $\sqrt{5}$-phase.

 \begin{figure}
 \centering
 \scalebox{0.40}{\includegraphics{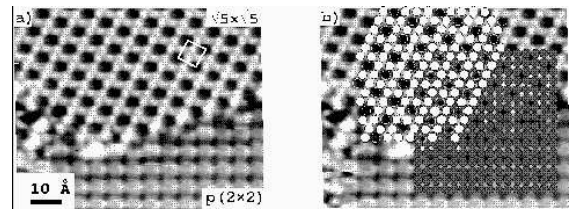}}
 \caption{a) STM image showing a domain boundary between the
 Pd(100)-$(\sqrt{5} \times \sqrt{5})R27^o$-O and the $p(2 \times
 2)$ phase. b) The same STM image but with a Pd(100) lattice
 superimposed (grey circles). This shows directly that in the
 $\sqrt{5}$-phase bright spots (assigned to Pd atoms, white
 circles) are shifted in neighboring rows by half a nearest
 neighbor distance, and that the dark spots (Pd atoms in hollow sites,
 transparent circles) coincide with hollow
 sites of the underlying Pd(100) substrate (Tunnel parameters:
 V=0.76 V, I=0.57 nA).}
 \label{fig4}
 \end{figure}

 Finally, we also performed STM measurements of the
 $\sqrt{5}$-phase, a corresponding image of which is shown in Fig.
 3a. As may directly be seen from the image, neighboring bright
 rows are shifted by half a nearest neighbor distance with respect
 to each other, in contrast to what would be expected from the
 PdO(001) geometry depicted in Fig. 3a. This finding (STM) together
 with the absence of a split in the calculated O $1s$ spectrum
 (HRCLS) and the low energetic stability (DFT) led us to conclude
 that the prevalent LEED $\sqrt{5}$-model is incompatible with the
 three methods employed in the present study.

 \subsubsection{Searching for a new model}

 To identify an alternative geometry of the $\sqrt{5}$-phase, we
 first further analyze our experimental STM data in order to reduce
 the vast phase space of possible structural models. The bottom
 right half of the STM image shown in Fig. 4a exhibits a domain of
 the coexisting $p(2 \times 2)$ on-surface adsorption phase, in
 which the oxygen overlayer simply occupies fourfold hollow sites
 of the underlying Pd(100) substrate
 \cite{kolthoff96,orent82,stuve84,chang88a,chang88b}. Based on this
 known geometry and the frequent finding that oxygen would appear
 as dark spots in the STM image, we construct the Pd(100) lattice and
 superimpose it on the experimental image, as shown in Fig. 4b.
 It then follows that the dark spots in the $\sqrt{5}$-phase are
 directly situated on top of the fourfold hollow sites of the
 Pd(100) lattice.

 Assuming that the bright protrusions in the STM image correspond
 to the geometric position of Pd atoms, we may further draw the Pd
 sublattice of the suspected surface oxide layer into the STM image
 as done in Fig. 4b. This way, a total of three Pd atoms per
 $\sqrt{5}$ unit-cell are found, forming a rather open interlaced
 ring-like layer, the structure of which doesn't resemble a
 bulk-like PdO planar nearest-neighbor environment at all. The
 latter would instead be obtained, if Pd atoms would also be
 present at the position of the large dark spots, yielding then a
 more compact layer with a total of four Pd atoms per $\sqrt{5}$
 unit-cell. As the dark spots are directly situated on top of the
 hollow substrate sites, a straightforward explanation why these
 latter Pd atoms do not show up in the STM images would e.g. be a
 large corrugation within the surface oxide overlayer, in which all
 Pd atoms over Pd(100) hollow sites are strongly relaxed inwards.

 While we may thus tentatively determine the positions of the Pd
 atoms on the basis of the STM images, the latter do not lead to
 any conclusions about the position and number of oxygen atoms in
 the $\sqrt{5}$ unit-cell. Concerning the O coverage, we can
 however resort to the HRCLS measurements. Calibrating the spectra,
 cf. Fig. 1, with the $p(2 \times 2)$ (0.25\,ML) and $c(2 \times 2)$
 (0.50\,ML) adsorbate structures known from the previous LEED work
 \cite{kolthoff96,orent82,stuve84,chang88a,chang88b}, a rough
 estimate of $\theta \sim 0.8$\,ML is obtained, which would correspond
 to four O atoms per $\sqrt{5}$ unit-cell.

 Using these experimental observations, we proceed to set up
 structural models that are compatible with the data discussed so
 far. Assuming the $\sqrt{5}$-phase to be some form of surface
 oxide on Pd(100), PdO-like overlayers seem a most appealing choice
 for a model. A systematic look at all possible low-index PdO
 planes turns up two orientations which exhibit Pd positions whose
 lateral arrangement would agree with that deduced from STM: PdO(100) and
 PdO(101), cf. Fig. 3b and 3c respectively. In
 contrast, the earlier LEED model consists essentially of a
 PdO(001) plane on Pd(100), which is \em not \em equivalent to
 PdO(100) due to the tetragonal unit-cell of PdO \cite{rogers71}.
 Hence, the LEED model features an orientation, which does not fit
 the STM data, as seen when comparing Fig. 3a with Fig. 4.

 \subsubsection{PdO(101)/Pd(100) as the new model}

 Having filtered out PdO(100) and PdO(101) as two possible candidates
 for a new structural model, we note that the two differ only in the
 vertical position of the oxygen atoms. Each structure contains four oxygen
 and four Pd atoms per $\sqrt{5}$ unit-cell, which nicely fits the
 experimental coverage estimate described in Section III.2 . Yet, PdO(100) has
 all four O atoms above the Pd layer, while PdO(101) has two up and
 two down, as can be seen in Figs. 3b and 3c respectively. To
 discern between the two orientations, we subjected both to a full
 structure optimization in our DFT calculations. Interestingly,
 this yields a significantly increased stability for both overlayer
 models: in both cases, the binding energy is more than 1\,eV per O
 atom higher than that of the previous LEED model, thus providing
 the final evidence for the incorrectness of the latter. More
 precisely, we find the PdO(101)/Pd(100) $\sqrt{5}$-geometry to be
 the most stable of the three structural models depicted in Fig. 3
 with a binding energy of $E_B = +1.35$\,eV\,/\,O atom (compared to
 $E_B$ = $-0.12$\,eV\,/\,O atom for the LEED model and $E_B$ = +0.90\,eV\,/\,O
 atom for the PdO(100) overlayer).

 \begin{figure}
 \centering
 \scalebox{0.34}{\includegraphics{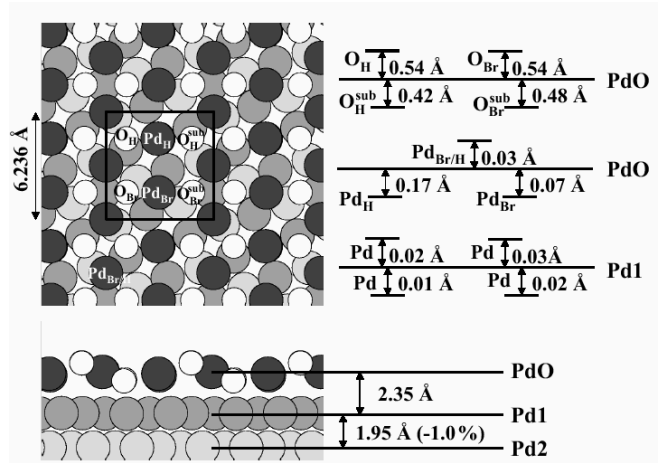}}
 \caption{Top- and side-view of the PdO(101)/Pd(100) model for the
 $\sqrt{5}$-phase on the basis of the DFT calculations. The
 rumpling of both the O and Pd atoms in the PdO overlayer and of
 the Pd atoms in the topmost substrate layer is given with respect
 to the center of mass of the respective layers. In the bottom
 right, also the average layer distances between these center of
 mass are indicated.} \label{fig5}
 \end{figure}

 Further checking on PdO(101)/Pd(100) as our new structural model,
 we also calculated a number of geometries keeping the positions of
 the Pd atoms in the overlayer, but testing different lateral
 positions for the O atoms. In particular, this involved
 geometries, where some oxygen atoms were located in bridge sites
 between the overlayer Pd atoms to produce differently coordinated
 O atoms at the surface that could then possibly also generate a
 split O $1s$ core-level spectrum. Yet, in all such combinatorial
 cases with O in hollow and bridge sites we obtained binding
 energies more than 0.2\,eV lower than for our PdO(101)/Pd(100)
 model. As second test series, we tried different registries of the
 PdO(101) overlayer on Pd(100), i.e. we laterally shifted the PdO
 overlayer around on the substrate. Again, this always resulted in
 a lower stability with respect to the structure shown in Fig. 3c.
 Finally, the dark holes seen in the STM image suggested that one of
 the four overlayer Pd atoms could be absent, namely the one
 over the fourfold hollow site, Pd${}_{\rm H}$ in Fig. 5. With bulk
 Pd as reservoir for the removed Pd atom, we also find such a model
 to be less stable. In conclusion, the DFT calculations therefore strongly
 favor the PdO(101)/Pd(100) structure for the $\sqrt{5}$-phase.

\begin{table}
\begin{center}
\begin{tabular}{|ll | c c c | c |} \hline
            &                   & Initial & Screening & Final & Experiment \\[0.1cm] \hline
I${}_{4f}$, & Pd${}_{\rm Br/H}$ & +0.79   & +0.10     & +0.89 &            \\
            &                   & +0.86   & +0.05     & +0.91 & \raisebox{1.5ex}[-1.5ex]{+1.30} \\ \hline
I${}_{2f}$, & Pd${}_{\rm Br}$   & +0.20   & +0.09     & +0.29 &            \\
            & Pd${}_{\rm H}$    & +0.18   & +0.21     & +0.39 & \raisebox{1.5ex}[-1.5ex]{+0.38} \\ \hline
II,         & Pd1               & -0.31   & +0.18     & -0.13 &            \\
            &                   & -0.28   & +0.19     & -0.09 &            \\
            &                   & -0.26   & +0.15     & -0.11 & -0.32      \\
            &                   & +0.01   & -0.05     & -0.04 &            \\
            &                   & +0.16   & -0.11     & +0.05 & \\ \hline
\end{tabular}
\end{center}
\caption{Calculated and measured Pd $3d$ surface core-level shifts
for the PdO(101)/Pd(100) model in eV. The computed values are
separated into initial-state and screening contribution, yielding
the total final-state shift that can be compared to experiment.
See Figs. 2 and 5 for the notation to describe the various first
(I) and second (II) layer atoms.}
\end{table}

 To see whether this new model is also compatible with the presented
 experimental HRCLS data, we show its computed final-state SCLS in
 Fig. 2 and list all its Pd $3d$ shifts in Table I. Comparing with the shifts
 obtained for the LEED model and for PdO(100)/Pd(100), cf. Figs. 3a and 3b,
 PdO(101) is the only model that exhibits an appreciably split O$1s$ core-level
 spectrum, due to the presence of both on- and sub-surface O in the geometry,
 cf. Fig. 2. The obtained Pd $3d$ shifts of +0.9\,eV and +0.4\,eV due to fourfold and
 twofold oxygen coordinated Pd atoms in the PdO(101) overlayer, Pd${}_{\rm
 Br/H}$ and Pd${}_{\rm Br}$/Pd${}_{\rm H}$ in Fig. 5, compare reasonably with
 the two experimental peaks that had already been assigned to differently
 coordinated Pd atoms on the basis of initial-state arguments. The remaining
 experimentally resolved peak with a small negative shift had similarly been
 attributed to the top Pd substrate atoms at the interface, which in the
 calculations exhibit almost vanishing SCLSs (Pd1 in Fig. 5).

 Of course, the very structure of the PdO(101) overlayer with an equal
 amount of on- and sub-surface oxygen atoms renders the measured
 splitted O $1s$ spectrum immediately plausible. In fact, the
 significant rumpling together with the different sub-surface O
 coordination to the underlying substrate, cf. Fig. 5, yield even
 slightly different shifts for the two atoms of each oxygen species
 present in the $\sqrt{5}$ unit-cell, cf. Fig. 2. Averaging the
 contributions within each group, we obtain a computed initial (final) state
 shift of 0.55 eV (0.49 eV) between the O $1s$ peaks due to on- and sub-surface
 oxygen atoms, in reasonable agreement with the measured value of 0.75 eV.

 \subsubsection{Compatibility with existing LEED data}

 So far, we have shown that our new model is superior to the model
 of Saidy {\em et al.} \cite{vu94,saidy01} with respect to all
 three techniques employed in the present work (HRCLS, STM, DFT).
 On the other hand, the model of Saidy {\em et al.} has strong
 backing from quantitative LEED. Hence, a final verification of our
 model would be to establish its viability also by this method.
 To accomplish this goal, we performed a rather restricted set of
 LEED $I(E)$ calculations for our structural model, comparing them
 to the very set of experimental LEED data published by Saidy {\em
 et al.}  \cite{saidy01} (scanned and digitized from their Fig. 2).
 From this we may judge whether or not our geometry can yield LEED
 $I(E)$ spectra on par or superior to those of Saidy {\em et al.}
 Our quantitative LEED calculations utilized the TensErLEED program
 package \cite{Blu01}, employing 10 fully relativistic phase shifts
 and a first-principles, energy-dependent real part of the inner
 potential, both generated for the surface geometry of Fig.
 \ref{fig5} using Rundgren's phase shift program package
 \cite{Run01}. Where the Tensor LEED method \cite{Rou86,Rou92} was
 employed, care was taken to ensure the full-dynamic
 reproducibility of the results in the final step of the
 calculation. All non-structural parameters of the calculation were
 kept fixed at the values chosen by Saidy {\em et al.}

 As a first step, we simply used the exact optimized geometry of
 the DFT-GGA calculations as input to the full-dynamic part of the
 TensErLEED code. Already this produced $I(E)$ curves in reassuring
 visual agreement with the scanned experimental spectra -- i.e.,
 all major spectral features could be reproduced. Still, shifts
 between individual peaks and overall shape difference only allowed
 for an average Pendry R-factor \cite{Pen80} $R_{\rm P}$ = 0.51
 between calculated and scanned $I(E)$. Hence, in a second step, we
 used the Tensor LEED method to relax all vertical positions in the
 PdO(101) layer, as well as the topmost two Pd(100) substrate
 layers below. The result of this is a clear drop of the best-fit
 Pendry R-factor to $R_{\rm P}$ = 0.28, shared by both integer
 ($R_{\rm int}$ = 0.29) and fractional ($R_{\rm frac}$ = 0.28)
 beams on average. The improvement is mainly due to overall
 slightly expanded distances between the individual layers compared
 to the DFT-GGA result. Moreover, a significant buckling is found
 in the second substrate layer, which was not relaxed in DFT. Of
 course, some differences of this kind must be expected already
 because no lateral or non-structural degrees of freedom were
 adjusted in the LEED fit. While the latter parameters may well
 account also for the remaining discrepancies between calculated
 and experimental $I(E)$ curves, the main goal of our LEED
 calculations has clearly been achieved: Already a very limited
 structural refinement of PdO(101)/Pd(100) produces
 experiment-theory agreement at a level which is even slightly
 improved compared to that presented by Saidy {\em et al.} ($R_{\rm
 P}$ = 0.306)\cite{saidy01} in their analysis. The consistency of
 our model with all available experimental data is thus established.

 \subsubsection{Strained PdO(101)/Pd(100)}

 The new structural model for the $\sqrt{5}$-phase is essentially a
 strained and rumpled PdO(101) film on top of Pd(100). The PdO(101)
 in-plane lattice constant is almost equal to that of a $\sqrt{5}$
 unit-cell on Pd(100), with the unit surface area of the
 commensurable film found here smaller by only 1.4\% than for
 unstrained PdO(101). 
 On the other hand, we compute a rather strong
 coupling of 100 meV/{\AA}${}^2$ of the laterally compressed PdO
 overlayer to the underlying Pd(100) substrate, rationalizing the
 formation of a commensurable surface oxide structure. This strong coupling
 also helps to stabilize the particular PdO(101)
 orientation, which is experimentally not found to be a preferred
 growth direction of PdO crystallites \cite{mcbride91}. Our
 calculations show that the stoichiometric termination of bulk
 PdO(101)-($1 \times 1$) suggested in the $\sqrt{5}$-film, i.e. the
 one terminated by O atoms as shown in Fig. 5, is in fact
 considerably more stable than the two other ways of truncating PdO
 in (101) direction, 57 meV/{\AA}${}^2$ compared to 134
 meV/{\AA}${}^2$ (also O terminated) and 128 meV/{\AA}${}^2$ (Pd
 terminated) \cite{rogal02}. Interestingly, the bulk PdO(100)
 orientation shown in Fig. 3b exhibits even a significantly lower
 surface energy (33 meV/{\AA}${}^2$), while this orientation is in
 the commensurable thin film geometry discussed here energetically
 not as favorable as the PdO(101) $\sqrt{5}$-model. Evidently, the
 presence of oxygen at the oxide/metal interface yields a stronger
 coupling to the underlying substrate and is ultimately responsible
 for the higher stability of the PdO(101)/Pd(100) surface oxide
 geometry.

 This example of the stabilisation of a higher energy crystal face
 in thin oxide films due to strong interfacial coupling to the
 substrate adds another interesting aspect to the new physics found
 recently in studies concerning oxide formation at TM surfaces.
 Among other findings, the formation of incommensurable domains of
 low energy oxide faces has been reported for ruthenium single
 crystals \cite{over00,kim01}, delineating the opposite case to the
 results reported here, i.e. when the oxide orientation is more
 important than a good coupling to the underlying substrate.
 Apparently, the lower thermal stability of palladium oxides
 compared to RuO${}_2$ increases the importance of the oxide/metal
 interface. This is further supported by the surface oxide
 structure just found on Pd(111), which does not resemble any PdO
 bulk orientation at all \cite{lundgren02}.

 Experimentally, oxide thicknesses below about 20 {\AA} have been
 found in all of these cases, indicating either a slow growth kinetics
 once the thin films have formed or a thermodynamic hindrance to
 form thick bulk oxides. This could be of interest in oxidation
 catalysis, where such oxide patches forming on TM surfaces in the
 reactive environment are now discussed as the actually active material
 \cite{over00,hendriksen02,stampfl02,reuter03}. If a continued
 growth of these oxide films is not possible, so that their structure
 always remains significantly affected by the interfacial coupling, they may
 exhibit catalytic behavior which is non-scalable from corresponding
 bulk oxide crystallites - in other words, truly nano-catalytic properties.
 Even when the oxide growth is not limited, the structure of the initially
 formed oxide film will be crucial, setting the stage for the
 ensuing oxidation process. For thicker films, interfacial coupling
 will be progressively less influential, so that an initially
 stabilized higher energy oxide orientation as found in the present
 work should eventually become liable to faceting. The
 corresponding three-dimensional cluster growth has indeed been
 observed for the continued oxidation of both Pd(100) and Pd(111)
 \cite{zheng02,zheng00}.

 \section{Summary}

 In conclusion, we have shown that the prevalent structural model
 for the Pd(100)-$(\sqrt{5} \times \sqrt{5})R27^o$-O surface oxide
 can not be reconciled with neither the experimental nor the
 theoretical methods employed in the present study: Its surface
 symmetry does not fit to the one observed by STM, and the
 calculated HRCLS for this structure do not show the appreciable
 splitting of the O $1s$ spectrum observed experimentally. In
 addition DFT calculations give only a very low energetic
 stability and a relaxed geometry that does no longer exhibit the
 significant rumpling originally introduced to match the measured
 LEED $I(E)$ curves.

 Based on the present experimental data we reanalyze the
 $\sqrt{5}$-phase and suggest an alternative structural model: a
 strained PdO(101) layer on Pd(100). This arrangement is
 energetically much more stable in our DFT calculations. Its
 computed final-state SCLSs agree well with all HRCLS measurements,
 linking the large splitting of the O $1s$ spectrum to the presence
 of oxygen both at the surface and at the oxide/metal interface.
 Already a very restricted set of LEED intensity calculations
 establishes the compatibility of this structure also with the
 previously published LEED intensity data.

 The PdO(101) orientation, which is experimentally not found to be
 a preferred PdO growth direction, is stabilized by the strong
 coupling to the underlying substrate in the present thin film
 limit. In comparison to the ensuing three-dimensional cluster
 growth during continued oxidation, the $\sqrt{5}$-phase is
 therefore likely to display different physico-chemical properties,
 which might be of interest or relevance to high-pressure
 applications like catalysis.

 \section{Acknowledgements}
 We are thankful for partial support by the DFG priority program
 "Realkatalyse". The support from the MAXLAB staff and financial
 support from the Swedish Research Council is gratefully
 acknowledged. Stimulating discussions with Georg Kresse are also
 gratefully acknowledged.


\begin{references}

 \bibitem{schmalz91}
 A. Schmalz, S. Aminpirooz, L. Becker, J. Haase, J. Neugebauer,
 M. Scheffler, D.R. Batchelor, D.L. Adams, and E. B\o gh,
 Phys. Rev. Lett. {\bf 67}, 2163 (1991).

 \bibitem{burchhardt95}
 J. Burchhardt, M.M. Nielsen, D.L. Adams, E. Lundgren,
 J.N. Andersen, C. Stampfl, M. Scheffler, A. Schmalz,
 S. Aminpirooz, and J. Haase,
 Phys. Rev. Lett. {\bf 74}, 1617 (1995).

 \bibitem{stampfl96}
 C. Stampfl, S. Schwegmann, H. Over, M. Scheffler, and G. Ertl,
 Phys. Rev. Lett. {\bf 77}, 3371 (1996).

 \bibitem{lee00}
 S.-H. Lee, W. Moritz, and M. Scheffler,
 Phys. Rev. Lett. {\bf 85}, 3890 (2000).

 \bibitem{lundgren02}
 E. Lundgren, G. Kresse, C. Klein, M. Borg, J.N. Andersen, M. De
 Santis, Y. Gauthier, C. Konvicka, M. Schmid, and P. Varga, Phys.
 Rev. Lett. {\bf 88}, 246103 (2002).

 \bibitem{zheng02}
 G. Zheng and E.I. Altman, Surf. Sci. {\bf 504}, 253 (2002).

 \bibitem{vu94}
 D.T. Vu, K.A.R. Mitchell, O.L. Warren, and P.A. Thiel, Surf. Sci.
 {\bf 318}, 129 (1994).

 \bibitem{saidy01}
 M. Saidy, O.L. Warren, P.A. Thiel, and K.A.R. Mitchell, Surf. Sci.
 {\bf 494}, L799 (2001).

 \bibitem{mcbride91}
 J. McBride, K. Hass, and W. Weber, Phys. Rev. B {\bf 44}, 5016
 (1991).

 \bibitem{nyholm01}
 R. Nyholm {\em et al.}, Nucl. Instr. and Meth. A {\bf 467}, 520 (2001).

 \bibitem{blaha99}
 P. Blaha, K. Schwarz, and J. Luitz, {\bf WIEN97}, \emph{A Full
 Potential Linearized Augmented Plane Wave Package for Calculating
 Crystal Properties}, Karlheinz Schwarz, Techn. Universit\"at Wien,
 Austria, (1999). ISBN 3-9501031-0-4.

 \bibitem{kohler96}
 B. Kohler, S. Wilke, M. Scheffler, R. Kouba, and C.
 Ambrosch-Draxl, Comput. Phys. Commun. {\bf 94}, 31 (1996).

 \bibitem{petersen00}
 M. Petersen, F. Wagner, L. Hufnagel, M. Scheffler, P. Blaha, and
 K. Schwarz, Comp. Phys. Commun. {\bf 126}, 294 (2000).

 \bibitem{perdew96}
 J.P. Perdew, S. Burke, and M. Ernzerhof, Phys. Rev. Lett. {\bf
 77}, 3865 (1996).

 \bibitem{kolthoff96}
 D. Kolthoff, D. J\"urgens, C. Schwennicke, and H. Pfn\"ur, Surf.
 Sci. {\bf 365}, 374 (1996).

 \bibitem{pirovano99}
 M.V. Ganduglia-Pirovano and M. Scheffler, Phys. Rev. B {\bf 59},
 15533 (1999).

 \bibitem{li03}
 W.X. Li, C. Stampfl, and M. Scheffler, Phys. Rev. B {\bf 67}, 045408 (2003).

 \bibitem{CRC95}
 {\em CRC Handbook of Chemistry and Physics}, CRC press (Boca Raton, FL,
 1995).

 \bibitem{spanjaard85}
 D. Spanjaard, C. Guillot, M.C. Desjonqueres, G. Treglia, and J.
 Lecante, Surf. Sci. Rep. {\bf 5}, 1 (1985); W. F. Egelhoff, Surf.
 Sci. Rep. {\bf 6}, 253 (1987).

 \bibitem{lizzit01}
 S. Lizzit, A. Baraldi, A. Groso, K. Reuter, M.V.
 Ganduglia-Pirovano, C. Stampfl, M. Scheffler, M. Stichler, C.
 Keller, W. Wurth, and D. Menzel, Phys. Rev. B {\bf 63}, 205419
 (2001).

 \bibitem{orent82}
 T.W. Orent and S.D. Bader, Surf. Sci. {\bf 115}, 323 (1982).

 \bibitem{stuve84}
 E.M. Stuve, R.J. Madix, and C.R. Brundle, Surf. Sci. {\bf 146},
 155 (1984).

 \bibitem{chang88a}
 S.-L. Chang and P.A. Thiel, J. Chem. Phys. {\bf 88}, 2071 (1988).

 \bibitem{chang88b}
 S.-L. Chang, P.A. Thiel, and J.W. Evans, Surf. Sci. {\bf 205}, 117
 (1988).

 \bibitem{rogers71}
 D. Rogers, R. Shannon, and J. Gillson, J. Solid State Chem. {\bf
 3}, 314 (1971).

 \bibitem{Blu01}
 V. Blum and K. Heinz, Comp. Phys. Comm. {\bf 134}, 392 (2001).

 \bibitem{Run01}
 J. Rundgren, Phase shift package (private communication, 2001, and
 to be published).

 \bibitem{Rou86}
 P.J. Rous, J.B. Pendry, D.K. Saldin, K. Heinz, K. M\"uller, and N.
 Bickel, Phys. Rev. Lett. {\bf 57}, 2951 (1986).

 \bibitem{Rou92}
 P.J. Rous, Prog. Surf. Sci. {\bf 39}, 3 (1992).

 \bibitem{Pen80}
 J.B. Pendry, J. Phys. C {\bf 13}, 937 (1980).

 \bibitem{rogal02}
 J. Rogal, K. Reuter, and M. Scheffler, (to be published).

 \bibitem{over00}

 H. Over, Y.D. Kim, A.P. Seitsonen, S. Wendt, E. Lundgren, M.
 Schmid, P. Varga, A. Morgante, and G. Ertl, Science {\bf 287},
 1474 (2000).

 \bibitem{kim01}
 Y.D. Kim, A.P. Seitsonen, and H. Over, J. Phys. Chem B {\bf 105},
 2205 (2001).

 \bibitem{hendriksen02}
 B.L.M. Hendriksen and J.W.M. Frenken, Phys. Rev. Lett. {\bf 89},
 046101 (2002).

 \bibitem{stampfl02}
 C. Stampfl, M.V. Ganduglia-Pirovano, K. Reuter, and M. Scheffler,
 Surf. Sci. {\bf 500}, 368 (2002).

 \bibitem{reuter03}
 K. Reuter and M. Scheffler, Phys. Rev. Lett. {\bf 90}, 046103 (2003).

 \bibitem{zheng00}
 G. Zheng and E.I. Altman, Surf. Sci. {\bf 462}, 151 (2000).

 \end{references}
 \end{document}